\documentclass{article}
\usepackage{times,epsf}
\input psfig

\newcommand{\ale}{\ \raisebox{-.3ex}{$\stackrel{<}{\scriptstyle \sim}$}\ }
\newcommand{\age}{\ \raisebox{-.3ex}{$\stackrel{>}{\scriptstyle \sim}$}\ }

\begin{document}

\title{Accretion disc dynamics in extreme mass ratio compact binaries}
\author{M. R. Truss\thanks{E-mail: m.r.truss@durham.ac.uk}\\
Department of Physics, Durham University, South Road, Durham, DH1 3LE}
\date{}

\maketitle

\label{firstpage}

\begin{abstract}
An analysis is presented of a numerical investigation of the dynamics and geometry of  accretion 
discs in binary systems with mass ratios $q = M_2 / M_1 < 0.1$, applicable to ultra-compact X-ray 
binaries, AM CVn stars and very short period cataclysmic variables. The steady-state geometry of
the disc in the binary reference frame is found to be quite different from that expected
at higher mass ratios. For $q \sim 0.1$, the disc takes on the usual elliptical shape, with the
major axis aligned perpendicular to the line of centres of the two stars. However, at smaller
mass ratios the elliptical gaseous orbits in the outer regions of the disc are rotated in the binary plane. The 
angle of rotation increases with gas temperature, but is found to vary inversely with $q$. At 
$q = 0.01$, the major axis of
these orbits is aligned almost parallel to the line of centres of the two stars. These effects may be responsible for the similar 
disc structure inferred from Doppler tomography of the AM CVn star GP Com (Morales-Rueda et al. 2003), which has 
$q = 0.02$. The steady-state geometry at low mass ratios is not predicted by an inviscid, restricted three-body 
model of gaseous orbits; it is related to the effects of tidal-viscous truncation of the disc near the Roche lobe 
boundary. Since the disc geometry can be inferred observationally for some systems, it is proposed that this 
may offer a useful diagnostic for the determination of mass ratios in ultra-compact binaries.
\end{abstract}

\section{Keywords}
accretion, accretion discs - binaries: close - novae, cataclysmic variables - instabilities

\section{Introduction}

\subsection{Background}
In an interacting binary system, the tidal torque on the accretion disc due to the orbital motion of the secondary star 
plays two major roles. Firstly, it acts as a sink for the angular momentum that is transported outwards by viscous 
processes, thereby truncating the disc. Secondly, resonances between the orbital motion of the secondary star and the 
orbital motion of gas in the disc can drive eccentricity growth, precession and spiral waves. Both these processes 
have a profound impact on the dynamics of the disc and its geometry.

The influence of tides on discs has been studied in great detail for interacting binary systems with mass ratios (defined in this 
paper as the ratio of the mass of the donor star to the mass of the accreting star) $q \age 0.1$. The work has been well
motivated by the wealth of cataclysmic variables and X-ray binaries that are covered by this range, all with orbital 
periods of order hours. However, there is a growing body of observational data from ultra-compact systems with orbital 
periods of the order tens of minutes. These very tight, double-degenerate binaries are often inferred to have extremely 
small mass ratios. They comprise very low-mass ($M_2 \leq 0.1 {\rm M_\odot}$), hydrogen-poor donor stars transferring 
mass onto either a white dwarf (the AM CVn stars), neutron star or black hole (the ultra-compact X-ray binaries). In a 
recent paper, Deloye, Bildsten \& Nelemans (2005) tabulated existing mass ratio determinations for several of the AM CVn systems; these range 
from $q = 0.087$ for AM CVn itself down to $q = 0.0125$ for CE-315. It is timely, therefore, to revisit the subject of tides 
with reference to these systems.

\subsection{Tides and resonances in accretion discs}
Paczy\'nski (1977) calculated orbits in the restricted three-body problem for mass ratios greater than $q \sim 0.03$, and associated 
the truncation radius of a pressure-free, inviscid disc with the radius of the largest periodic orbit which does not intersect 
any other periodic orbits. He further found that some large-radius orbits were unstable for mass ratios $q < 0.25$. 

Early numerical calculations of viscous discs confirmed that at typical cataclysmic variable mass ratios the disc takes on
a steady structure similar to the shape of the inviscid three-body orbits calculated by Paczy\'nski (1977). The disc was found to 
become slightly elongated perpendicular to the line of centres of the two stars (Whitehurst \& King 1991). Furthermore, 
Whitehurst \& King 
found that their numerical discs became unstable below a critical mass ratio $q _{\rm \,crit} \sim 0.25-0.33$. 
They realised the significance of the 3:1 resonance between the orbits of gas in the disc and the orbital motion of the 
secondary star as a driver for eccentricity generation. This came after previous simulations had shown that superhumps - 
photometric modulations in the light curve of low mass ratio dwarf novae - could be generated by an eccentric, precessing 
disc (Whitehurst 1988). Hirose \& Osaki (1990) parameterised the resonance condition, showing that the width of the resonance
is proportional to the reduced mass of the binary. It follows that the range of resonant radii narrows as $q$ decreases.

Lyubarskij, Postnov \& Prokhorov (1994) have confirmed that a circular disc with the standard alpha viscosity is viscously unstable to eccentric 
perturbations. More recently, Ogilvie (2001) used a viscoelastic theory for turbulent stress to show that the instability 
can be suppressed by the presence of a large enough relaxation time or bulk viscosity. The theory of tidal instability was 
refined by Lubow (1991), who highlighted the coupling between the tidal potential and
modes excited in the disc on contact with the resonance. Any small existing eccentricity is swiftly amplified, with the
growth rate of the eccentricity proportional to the strength of the two-armed spiral mode. This model, and its application
to dwarf nova outbursts has been confirmed many times by numerical simulations (Kunze, Speith \& Riffert 1997; Murray 1996, 1998; Truss, Murray \& Wynn 2001; Truss 2005).

It is not obvious how the tidal instability
manifests itself in systems with very small mass ratios. The growth rate of the instability is proportional to $q^2$
(Lubow 1991) and hence will be extremely slow, while the range of resonant radii is also extremely narrow (Hirose \& Osaki 1990).
Simpson \& Wood (1998) found precessing, eccentric discs in numerical simulations of binaries with mass ratios $q = 0.05, 0.075$ and 
$0.1$, but {\em not} in binaries with $q = 0.025$. This suggests that there may be a lower limit below which the instability
does not operate, although many more high resolution simulations would be required to verify this.

Recently, there has been renewed interest in the role of the 2:1 resonance at very small mass ratios. Osaki \& Meyer (2002)
have suggested that the orbital-period photometric humps that appear in the outburst light curves of the low mass-ratio 
cataclysmic variable WZ Sagittae are caused by the 2:1 resonance. Numerical simulations by Kunze \& Speith (2005) have
supported this hypothesis.

\subsection{Outline of paper}

This paper addresses the question of the steady-state structure of an accretion disc in an interacting binary with $q \leq 0.1$.
It is shown that the structure of the disc at these mass ratios does not agree with the prediction of an inviscid, three-body 
model. Rather than taking on a structure that is elongated along an axis perpendicular to the line of centres of the two 
stars, the disc takes on an elliptical form, offset at an angle to this axis. The angle increases with disc temperature and 
increases with decreasing $q$. This effect is related to the tidal truncation of the disc
near the Roche lobe. It is not a result of the tidal instability: the analysis 
presented here refers to discs that have not yet become tidally unstable.

In the next section, the inviscid, restricted three-body model is revisited briefly for systems with low mass ratios. This is
followed by a discussion of the results of a series of hydrodynamic simulations of viscous discs. The steady structure of an
accretion disc is determined for different mass ratios, and the effect on this structure of different gas temperatures and
mass transfer rates is explored.

\section{The size of an accretion disc}

\begin{figure}
\psfig{file=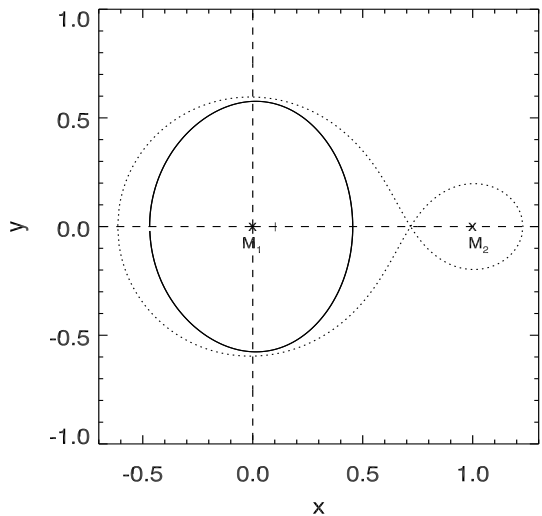,width=8cm,height=7.2cm}
\caption{Coordinate system used in the orbit calculations. For reference, the orbit shown here is the last 
non-intersecting orbit for $q = 0.1$, which has an eccentricity $e = 0.54$. The Roche lobe is also plotted for this mass ratio.}
\label{fig1}
\end{figure}
In the limit of vanishingly small pressure and viscosity, the streamlines of an accretion disc in a binary star potential can be
described by periodic orbits in the restricted three-body problem (Lubow \& Shu 1975). Paczy\'nski (1977)identified the maximum radius
of such a disc with that of the largest simple non-intersecting periodic orbit. Following this methodology, the maximum
radius is found by integration of periodic test-particle orbits in the Roche potential. Figure \ref{fig1} shows the system of 
coordinates that are adopted. The mass ratio is defined in the modern convention as
\begin{equation}
	q = \frac{M_2}{M_1}
\end{equation}
where the accretion disc is centred on the primary star of mass $M_1$. This definition is the inverse of that used in 
Paczy\'nski (1977). The reduced mass is
\begin{equation}
	\mu = \frac{q}{1+q},
\end{equation}
which here is the distance from the centre of the primary star to the centre of mass (denoted by a short vertical line in
Figure \ref{fig1}).  Periodic solutions are found by integrating orbits with a given value of the Jacobi constant, 
and demanding that after one orbit, $x = x(0)$ and $\dot x  = 0$. The condition $\dot x = 0$ ensures that the orbits are 
symmetric with respect to the x-axis. Furthermore, the stability of the periodic orbits can be investigated following the 
method of H\'enon  (1965) and Piotrowski \& Ziolkowski (1970).
\begin{figure}
\begin{center}
$\begin{array}{c@{\hspace{-0.25cm}}c}
\epsfxsize=4.25cm
\epsfysize=4.0cm
\epsffile{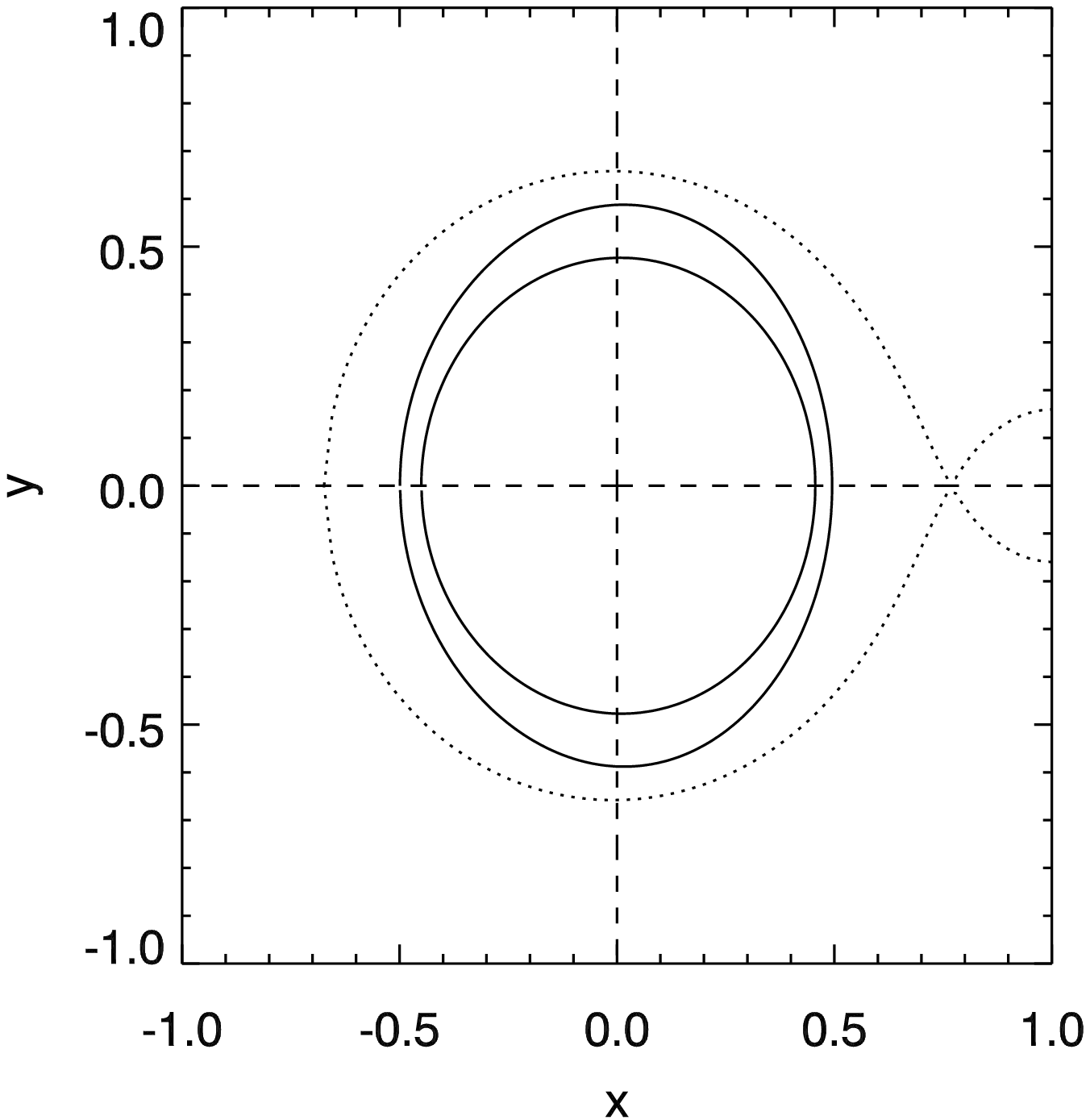} &
	\epsfxsize=4.25cm
	\epsfysize=4.0cm
	\epsffile{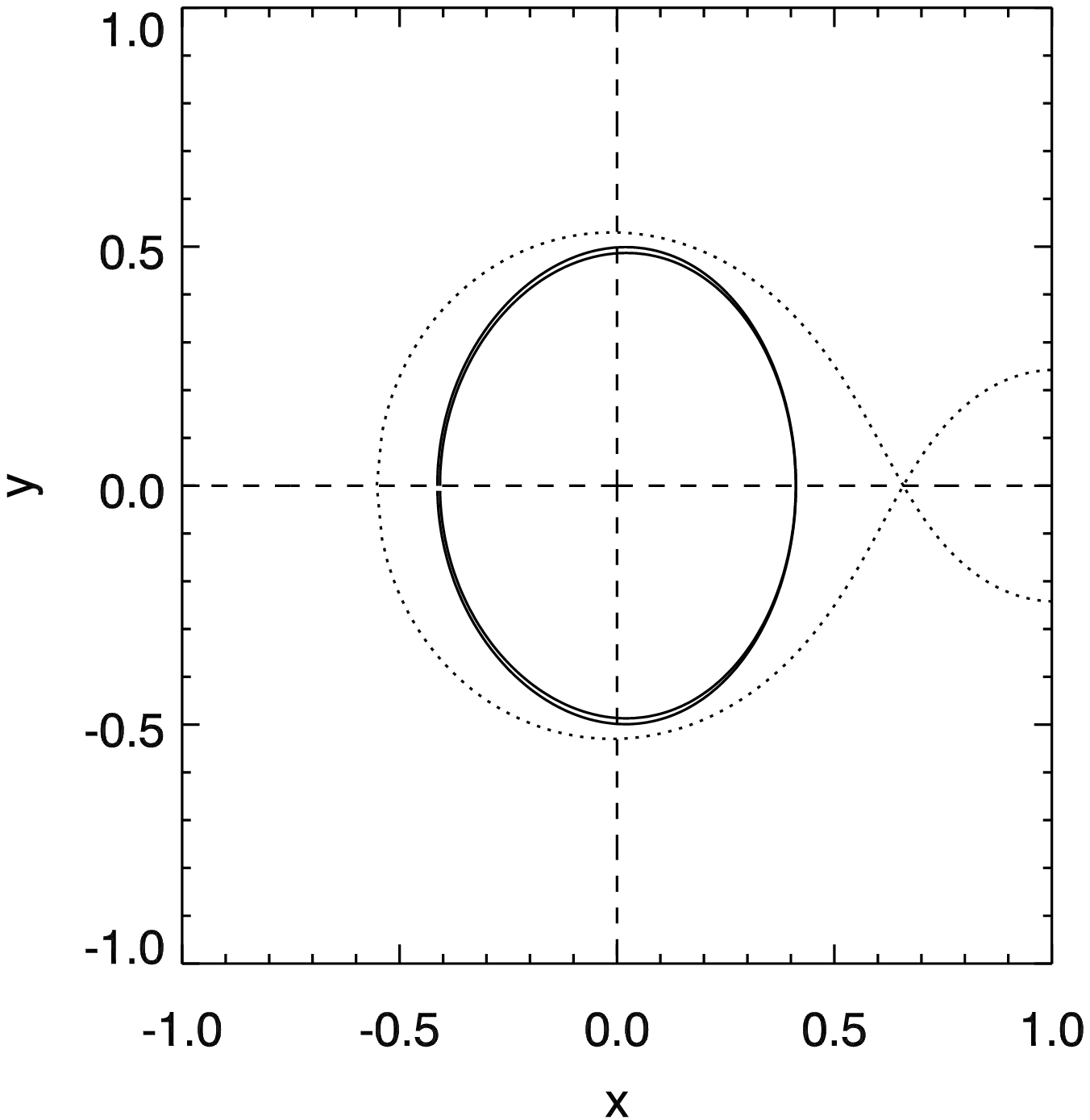} \\
\mbox{\bf (a)} & \mbox{\bf (b)}
\end{array}$
\end{center}
\caption{Last stable and last non-intersecting restricted three-body periodic orbits for (a) $q=0.05$ and (b) $q=0.2$. The 
two orbits are very close at $q=0.2$ as this mass ratio is close to the stability boundary near $q=0.25$. The Roche lobe for 
both stars is shown in each case. The origin of the coordinate system is the centre of the primary star, with the centre of the
secondary star at $x=1.0$. Note the high degree of symmetry in the orbits either side of the x-axis.}
\label{orbs}
\end{figure}
The resulting radii for the last non-intersecting  and last stable orbits agree very closely with those tabulated by Paczy\'nski (1977)
and Hirose \& Osaki (1990), so it is not necessary to reproduce them here. However, it is instructive to show the characteristic shape of 
these orbits for comparison with later viscous simulations. Figure \ref{orbs} shows the last non-intersecting and last stable
orbits for mass ratios $q=0.05$ and $q=0.2$. Their shape is typical of all the last non-intersecting orbits found down to 
$q=0.01$: prolate with a high degree of symmetry either side of the line of centres between the two stars (the x-axis in 
Figure \ref{fig1}). Two other factors were highlighted by this exercise. Firstly, the width of the unstable region of
orbits was found to decrease with mass ratio, agreeing with the results of Hirose \& Osaki (1990). Indeed, the range of 
unstable orbits becomes so small below $q=0.02$ that one needs a very accurate integrator to resolve the stability 
criterion at all. 

The second factor relates to the last non-intersecting orbit and is important for a discussion of the
truncation of an accretion disc. One can see immediately from Figures \ref{fig1} and \ref{orbs} that the maximum radius
of these orbits is close to the Roche lobe. This is encouraging, as one naturally expects the Roche lobe to be the absolute
maximum radius for retention of a disc, with material outside this radius lost from the system. However, we must not 
forget that these simple three-body orbits are only appropriate for the completely inviscid case. Real discs are not
inviscid; indeed, viscosity plays a crucial role in the transport of angular momentum and thereby in the truncation of an
accretion disc. It seems quite possible that for arbitrarily small $q$, the inviscid three-body model will predict no orbit
crossings inside the Roche lobe at all (again, a very accurate integration would be required to prove this beyond doubt).    
However, it is the presence of a viscosity that breaks the symmetry of the orbits about the line of centres, making the
orbits slightly aperiodic and generating the torque required to tidally truncate the disc. 

Therefore, the question of whether any of the results from an inviscid calculation can be applied to real discs is something 
that can only be tested under more physically reasonable conditions. In the next section I present hydrodynamic 
simulations of viscous accretion discs in order to assess the validity of the three-body approximation to more realistic 
conditions and to make a systematic investigation of the dynamical response of the disc to different mass injection rates 
and viscosities.

\section{Numerical hydrodynamics of accretion discs}

\subsection{Computational method}
The code is a refined version of the three dimensional smoothed particle hydrodynamics (SPH) accretion disc code 
described by Truss (2005) and references therein (see Monaghan (1992) for a complete review of the SPH technique). The 
refinement for this work is that the condition of absolute isothermality has been relaxed. While the scheme remains 
vertically isothermal, the sound speed varies with radius according to:
\begin{equation}
c_{\rm s} = c_{\rm 0} R^{-\frac{3}{8}}.
\end{equation}
where the quantity $c_{\rm 0}$ is expressed as a fraction of $a\Omega$, where $\Omega$ is the orbital angular velocity of 
the binary and $a$ is the binary separation. The radius $R$ is expressed as a fraction of $a$. The smoothing length is variable
according to the local density and has a maximum value $0.007a$.

\begin{figure*}
\psfig{file=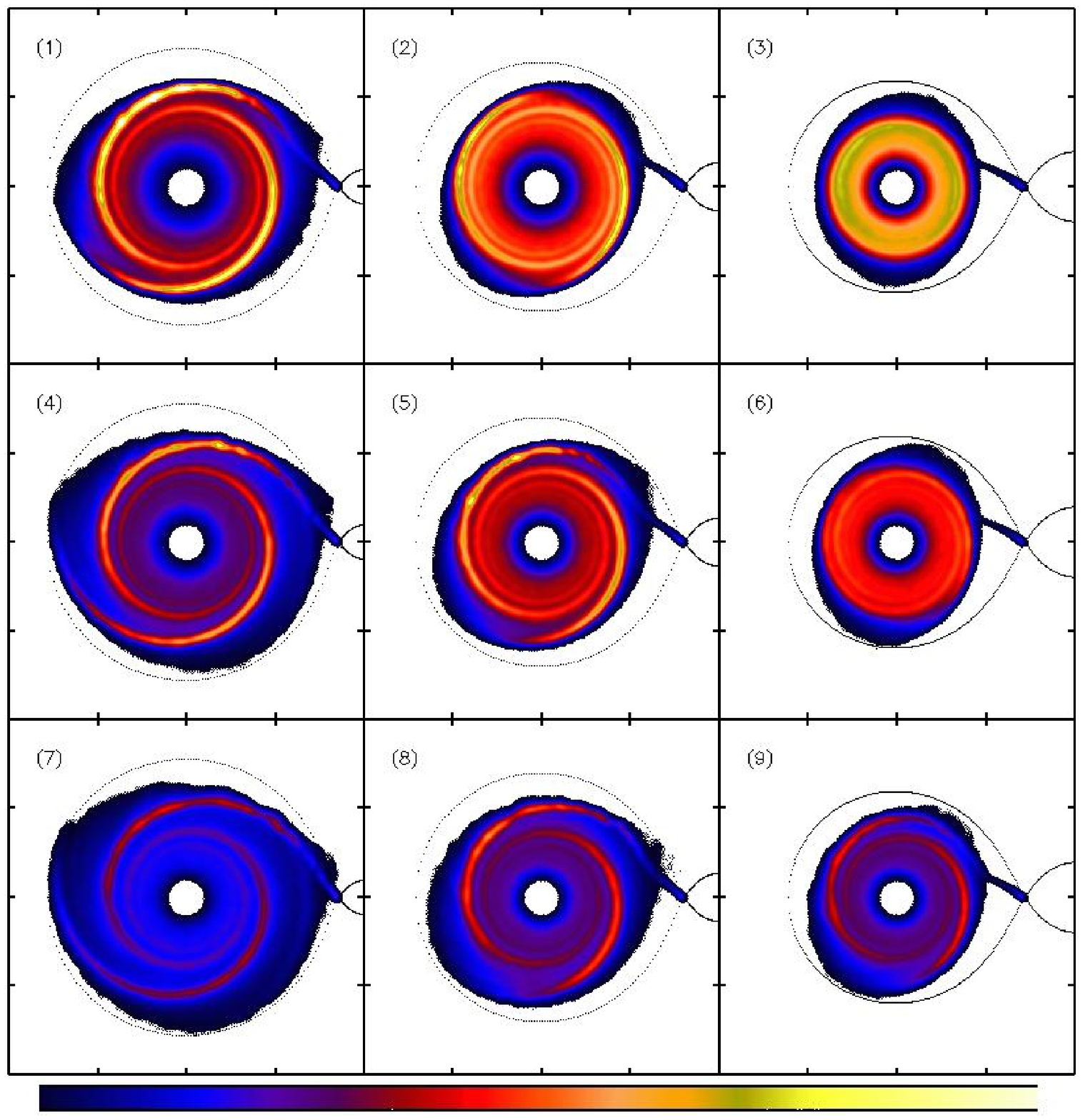,width=12cm}
\caption{Differences in the surface density structure of an accretion disc according to mass ratio and gas temperature. Each column 
shows the disc for different gas temperatures at a single mass ratio. Each row shows the variation in structure for 
different mass ratios at a single gas temperature. The columns from left to right show discs in binaries with $q = 0.01$, 
$q = 0.03$ and $q = 0.10$ respectively. The rows from top to bottom correspond to $c_{\rm 0} = 0.025$, $c_{\rm 0} = 0.035$ and 
$c_{\rm 0} = 0.050$ respectively.  Labels correspond to the run numbers given in Table \ref{tab1}. The colour scale is normalised 
such that it is identical for all the discs in Figures \ref{bigone} and \ref{bigtwo}. Each box is of side $a$ (one binary separation), 
with the origin at the centre of the primary star. The Roche lobe is plotted for clarity in each case.}
\label{bigone}
\end{figure*}

\begin{figure*}
\psfig{file=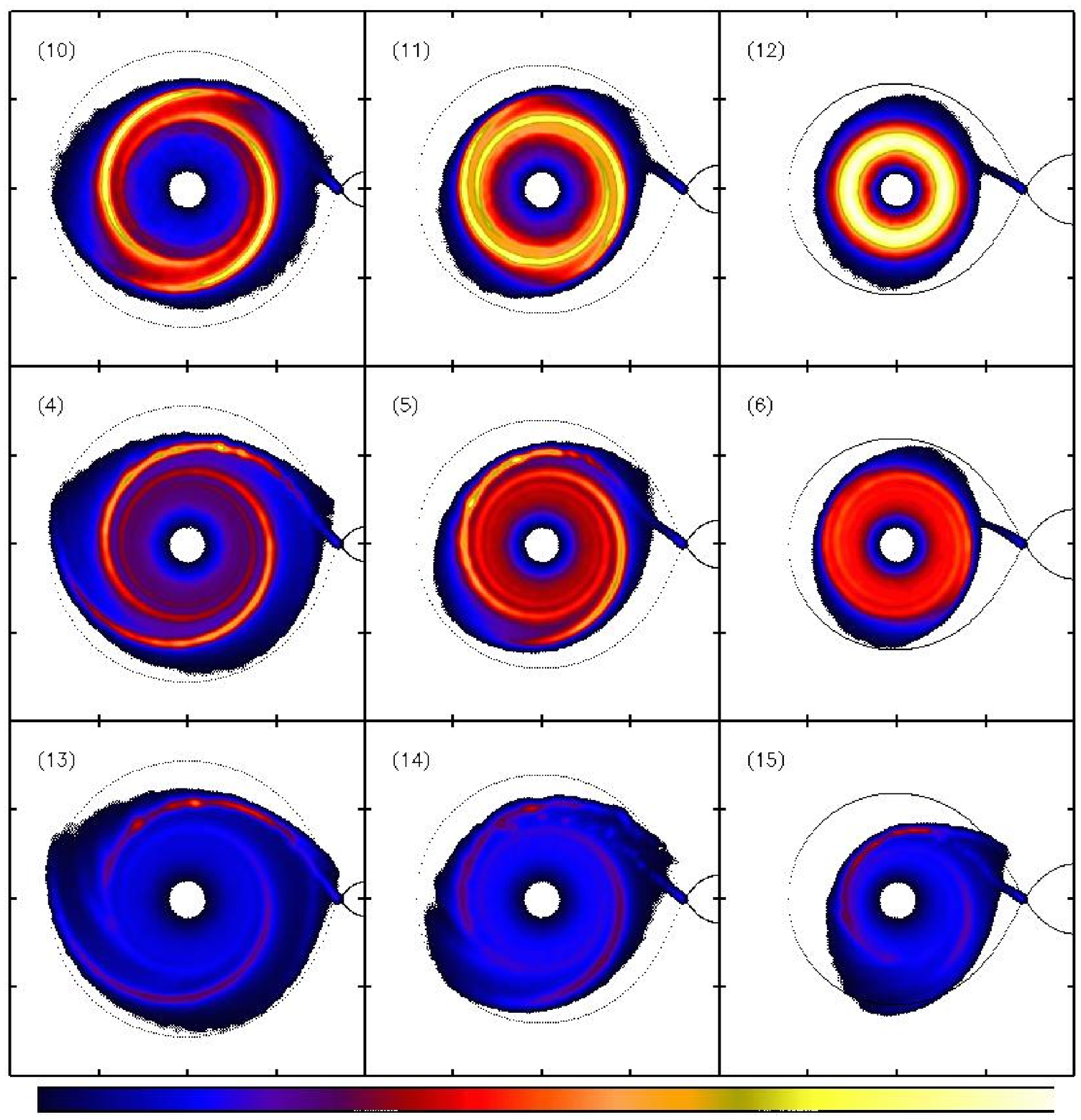,width=12cm}
\caption{Differences in the steady structure of an accretion disc according to mass ratio and viscosity parameter. Each column 
shows the disc for different gas temperatures at a single mass ratio. Each row shows the variation in structure for 
different mass ratios at a single gas temperature. The columns from left to right show discs in binaries with $q = 0.01$, 
$q = 0.03$ and $q = 0.10$ respectively. The rows from top to bottom correspond to $\alpha = 0.02$, $\alpha = 0.1$ and $\alpha = 0.5$
respectively. Labels correspond to the run numbers given in Table \ref{tab1}: the $\alpha = 0.1$ runs are identical to those shown
in Figure \ref{bigone}. Each box is of side $a$ (one binary separation), with the origin at the centre of the primary star. The Roche 
lobe is plotted for clarity in each case.}
\label{bigtwo}
\end{figure*}

\begin{table*}
\begin{tabular}{|c|c|c|c|c|c|c|c|} \hline
Run & $q$ & $c_{\rm 0}\,(a\Omega)$ & $\alpha$ & $-\dot M_2\,({\rm M_\odot yr^{-1}})$ & $N$ & $\theta_{\rm max} (degrees)$ & 
$e_{\rm max}$ \\ \hline
1 & 0.01 & 0.025 & 0.10 & $10^{-11}$ & 169634 & 77 & 0.57  \\
2 & 0.03 & 0.025 & 0.10 & $10^{-11}$ & 158992 & 36 & 0.55  \\
3 & 0.10 & 0.025 & 0.10 & $10^{-11}$ & 128414 & 7 & 0.59 \\
4 & 0.01 & 0.035 & 0.10 & $10^{-11}$ & 146443 & - & - \\
5 & 0.03 & 0.035 & 0.10 & $10^{-11}$ & 139830 & 55 & 0.60 \\
6 & 0.10 & 0.035 & 0.10 & $10^{-11}$ & 109584 & 17 & 0.55 \\
7 & 0.01 & 0.050 & 0.10 & $10^{-11}$ & 105129 & $>$90 & - \\
8 & 0.03 & 0.050 & 0.10 & $10^{-11}$ & 107740 & 63 & 0.57 \\
9 & 0.10 & 0.050 & 0.10 & $10^{-11}$ & 80838 & 29 & 0.60 \\
10 & 0.01 & 0.035 & 0.02 & $10^{-11}$ & 162678 & 86 & 0.52 \\
11 & 0.03 & 0.035 & 0.02 & $10^{-11}$ & 163167 & 41 & 0.58 \\
12 & 0.10 & 0.035 & 0.02 & $10^{-11}$ & 137284 & 16 & 0.58 \\
13 & 0.01 & 0.035 & 0.50 & $10^{-11}$ & 87053 & $>$90 & - \\
14 & 0.03 & 0.035 & 0.50 & $10^{-11}$ & 77120 & 63 & 0.57 \\
15 & 0.10 & 0.035 & 0.50 & $10^{-11}$ & 50551 & - & - \\
16 & 0.03 & 0.035 & 0.10 & $10^{-13}$ & 139825 & 57 & 0.62 \\
17 & 0.03 & 0.035 & 0.10 & $10^{-9}$ & 139812 & 52 & 0.58 \\ \hline
\end{tabular}
\caption{Data for each simulation. $N$ is the number of particles at the instant of each snapshot in Figures \ref{bigone}, 
\ref{bigtwo} and \ref{mdots}. $\theta_{\rm max}$ is defined as the tilt angle of the major axis of the outermost disc orbit with 
respect to the y axis of these plots. $e_{\rm max}$ is the eccentricity of the outermost disc orbit. The disc in run 15 is no longer in 
a steady state and has started to precess.}
\label{tab1}
\end{table*}

\subsection{Structural dependence on mass ratio and gas viscosity}
A fiducial set of binary parameters was chosen to be representative of ultra-compact binary systems. They were:
$M_{\rm 1} = 0.5 \,{\rm M_\odot}$, $P_{\rm orb} = 50 \,{\rm minutes}$ and $-\dot M_2 = 10^{-11} \,{\rm M_\odot yr^{-1}}$. 
In the first set of simulations, a total of nine calculations were performed comprising three disc temperatures at each of three 
mass ratios, $q = 0.01, 0.03$ and $0.10$. The three
sound speed parameters used were $c_{\rm 0} = 0.025$, $c_{\rm 0} = 0.035$ and $c_{\rm 0} = 0.050$. For these nine 
calculations, the Shakura/Sunyaev viscosity parameter was set to $\alpha = 0.1$. The temperatures
at the outer edge of the disc that the sound speeds correspond to depend weakly on the binary parameters for each 
simulation, but as a guide if $q = 0.01$, $c_{\rm 0}=0.025$ corresponds to a mid-plane temperature of about 
$25,000\,{\rm  K}$ (for $\mu = 1$). A second set of calculations was performed to investigate the dependence of the disc 
structure on the Shakura/Sunyaev viscosity parameter. These calculations comprised six further runs with $\alpha = 0.02$ and
$\alpha = 0.5$ at each mass ratio, all with $c_{\rm 0} = 0.035$, to complete a similar $3\times3$ grid of calculations in $q-\alpha$ space.  
Table \ref{tab1} summarises the input parameters for these and the further simulations described in this paper, together with
information regarding the number of particles involved for each run. 

Each disc was built up in three dimensions from scratch, injecting SPH particles from the $L_1$ point at the correct angle 
to the line of centres prescribed by Lubow \& Shu (1975). The discs were allowed to evolve in this way and settle to a steady, 
circularised state in which the total eccentric mode strength had 
ceased to decline. This is an important criterion to satisfy, as during the initial stages of disc formation the total eccentric 
mode strength is high. The mode strength decreases as the disc material is circularised and spreads under the action of 
viscosity. After it reaches a minimum value, the disc will either continue to remain circular, or become eccentric if parts of 
it can access the 3:1 resonance. All but one of the discs described here are fully circularised, with eccentric mode strengths in the 
range $10^{-4} < S(1,0) < 5 \times 10^{-3}$ (the exception, run 15, is discussed below).  The time taken to achieve this fully 
circularised state corresponds to many tens of orbital 
periods, or many thousands of dynamical times of the disc. As a benchmark guide, the mode strength must reach several 
times $10^{-3}$ to produce even the tiniest amount of disc precession (Truss 2005). Therefore it can be stated with certainty
that these discs are well developed, but not tidally unstable. This does not preclude them from becoming tidally 
unstable given an indefinite amount of time to evolve; the growth rate of the instability is, as we have seen previously,
extremely slow.

Figure \ref{bigone} shows the results of the first nine simulations. It can be seen immediately that the steady shape of the disc
fixed in the binary frame is, in general, not that suggested by the shape of the restricted three-body orbits. As mass ratio
decreases, or as sound speed increases, the major axis of the elliptical orbits in the outer parts of the disc becomes rotated
in the x-y plane. The closest approximation to the shape predicted by the restricted three-body
orbits (and previous calculations of discs in binaries with larger mass ratios) is that shown for run 3 in Figure \ref{bigone}
($q = 0.10$ and $c_{\rm 0} = 0.025$). The major axis for this disc is aligned nearly perpendicular to the line-of-centres.
However, we can see that the disc in run 1, which has a similar temperature to that in run 3 but has $q = 0.01$, is quite 
different. The major axis of this disc is aligned nearly parallel to the line-of-centres.  

The rotation angle, $\theta_{\rm max}$, is defined as the tilt angle between the major axis of the outermost disc orbit and the line perpendicular
to the line-of-centres. The maximum eccentricity is defined here as
\begin{equation}
e_{\rm max} = \sqrt {\left( 1 - \frac{b^2}{a^2}\right)}
\end{equation}
where $a$ and $b$ are the semi-major and semi-minor axes of the outermost disc orbit. These quantities are summarised in Table
\ref{tab1}.

Further inspection of Figure \ref{bigone} reveals that the rotation angle of the orbits is correlated strongly with the distance 
from the disc edge to the $L_1$ point (or the length of the mass transfer stream). A hotter disc spreads further under the 
action of viscosity, while the smaller the mass ratio, the larger the fraction of the Roche lobe of the primary star that can be
populated by stable orbits. There is little variation in maximum eccentricity of the discs, with a mean value $<e_{\rm max}> = 0.58$
and a standard deviation $\sigma = 0.03$.

It is noteworthy that the spiral arms become less tightly wound for hotter discs as one would expect, but the same effect
is also seen for discs in binaries with smaller and smaller mass ratios. In all cases, the spiral arms are confined within the Roche
lobe of the primary star. The structure of the hot discs at $q = 0.01$ (runs 4 and 7) is complex, showing asymmetry in both the x- 
and y- directions. At this mass ratio and these
temperatures, the orbits are no longer elliptical, so no value of $e_{\rm max}$ can be ascribed to them. The mass transfer 
stream can be seen joining the near-side spiral arm very close to the $L_1$ point. Indeed, the disc extends out to the Roche lobe 
boundary across a wide range of azimuth in these discs, and there was a significant rate of mass loss from the disc edge in these 
simulations. A similar situation may occur in discs around weakly magnetised white dwarfs, where the outer edge of the disc is 
propelled outwards by the magnetic field. Matthews et al. (2006) have discussed the implications of this mass loss (in the case of a magnetised 
primary) for our understanding of binary star evolution.

Figure \ref{bigtwo} shows the results of the second set of simulations (runs 10-15), in which the gas viscosity parameter was
varied at each mass ratio. Runs 4-6 are reproduced in Figure \ref{bigtwo} to aid comparison. Again, the values derived for 
$\theta_{\rm max}$ and $e_{\rm max}$ are summarised in Table \ref{tab1}. The results are very similar to those in Figure 
\ref{bigone}, with the expected trend of increasing $\theta_{\rm max}$ with viscosity parameter. There is one notable exception,
however. Run 15 ($q$ = 0.10, $c_{\rm 0}$ = 0.035 and $\alpha = 0.5$) is no longer in a steady state and has started to precess. The
viscous spreading at this $\alpha$ is very rapid and the disc edge quickly encounters the 3:1 resonance. However, even though the
thermal characteristics of the discs in runs 13 and 14 are identical (they differ only in mass ratio), neither of those discs is close to 
finding a state of precession. This is a direct demonstration that the growth rate of the tidal instability is very slow at small mass
ratios: having encountered the resonance, these discs need much longer to start precessing. 

\subsection{Structural dependence on mass transfer rate}
A further parameter that is known to influence the shape of an accretion disc is the magnitude of the mass transfer rate
from the companion star. The stream is a constant source of low specific angular momentum material, and this can act to
circularise the gaseous orbits (Lubow 1994). It is important, therefore, to investigate this effect for the systems with 
$q \le 0.1$ that are under study here. 

\begin{figure}
\psfig{file=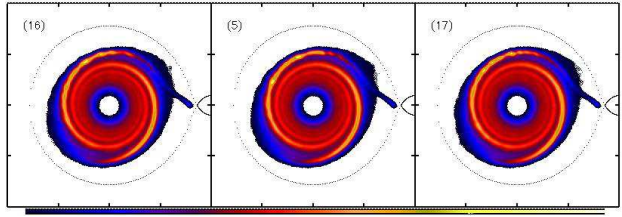,width=82mm}
\caption{The effect of varying the mass injection rate on disc structure for $q = 0.03$. From left to right, the mass transfer rates
are $\dot M = 10^{-13} \,{\rm M_\odot yr^{-1}}$, $\dot M = 10^{-11} \,{\rm M_\odot yr^{-1}}$ (run 5) and 
$\dot M = 10^{-9} \,{\rm M_\odot yr^{-1}}$ respectively. The particle mass in each simulation is scaled from the mass transfer 
rate; the surface densities in runs 16 and 17 are therefore 1/100th and 100 times those in run 5.}
\label{mdots}
\end{figure}

The simulation for $q = 0.03$, $c_{\rm 0} = 0.035$ and $\alpha = 0.1$ (run 5) was repeated using mass injection rates of 
$10^{-9}$ and $10^{-13} \,{\rm M_\odot yr^{-1}}$. The results are shown alongside the original 
$10^{-11} \,{\rm M_\odot yr^{-1}}$ run for comparison in Figure \ref{mdots}. While there is a weak trend for decreasing eccentricity 
and rotation angle with increasing mass transfer rate, the effect on the geometry of the steady disc at different mass transfer rates 
is not dramatic. It should be noted that this situation, in which each disc is built from scratch with a steady mass-transfer rate, is 
quite different from that in which a disc is subject to an instantaneous enhancement or interruption in mass transfer (see for 
example, Murray, Warner \& Wickramasinghe (2000)).

\begin{figure}
\psfig{file=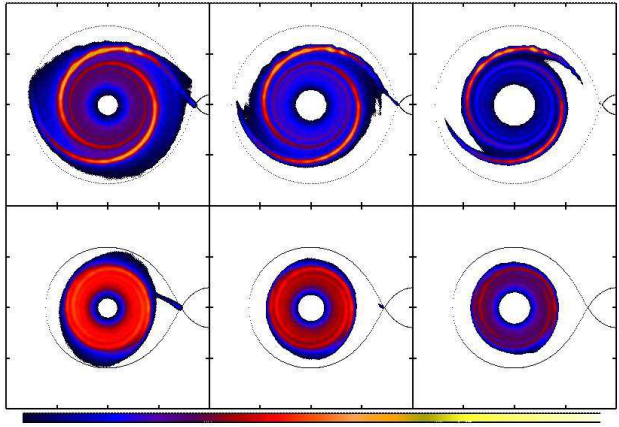,width=82mm}
\caption{The effect of neglecting low surface-density regions of the disc. The rows correspond to run 4 ($q$ = 0.01, top) and run 6
($q$ = 0.10, bottom). The left-hand panel shows the entire disc structure, while in the centre and right-hand panels, the lowest 10\%
and 20\% of the surface density have been removed respectively.}
\label{chop}
\end{figure}

\subsection{Other considerations}
The rotation angles and eccentricities derived in this work are determined using the outermost edge of the accretion disc, where
the surface density can be low. It is instructive, therefore, to investigate briefly the effect on the disc structure of neglecting the
lowest density regions. Figure \ref{chop} shows the effect of removing regions with less than 10\% and 20\% of the maximum 
surface density in representative simulations for $q$ = 0.01 (run 4) and $q$ = 0.10 (run 6). It is immediately clear that this does have
some effect on the inferred structure, especially at $q$ = 0.10. In this case, the rotation angle disappears and the eccentricity falls at
the 10\% level and the disc appears completely circular at the 20\% level. At $q$ = 0.01, a large rotation angle and eccentricity are
still measurable at the 10\% level, but things become less clear at the 20\% level. Here, the main body of the disc is circular apart
from the trailing ends of the two spiral arms.

\begin{figure}
\psfig{file=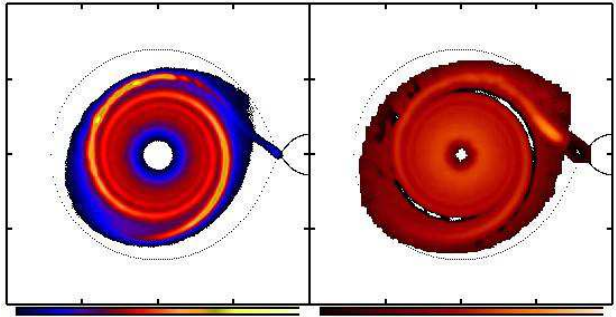,width=82mm}
\caption{A comparison of surface density structure (left) and energy dissipation rate (right) for run 5 ($q$ = 0.03). Both quantities 
yield consistent results for the angle $\theta_{\rm max}$ and the maximum eccentricity. High density regions do not 
necessarily correspond to regions of large emission: note the extended impact hot spot and the regions of very low emission just
behind the spiral arms.}
\label{comp}
\end{figure}

It is also important to examine the difference in the inferred structure if we map the viscous energy dissipation rate in the disc
rather than the surface density. This may offer a better comparison with observational tomographic techniques which map disc
emission. Figure \ref{comp} shows the two quantities mapped onto the disc for $q$ = 0.03 (run 5). While there is not necessarily a
strong correlation between regions of high emission and regions of high surface density - the extended stream impact hot-spot is a 
good example - the measured rotation angle and eccentricity are consistent for both quantities. Indeed, this is a general result for
all the simulations presented here. However, it should be noted when comparing with observations that Doppler tomograms are 
reconstructed from single emission lines and that these lines are produced in different regions of the accretion disc according to
the local conditions.

\section{Discussion}

It has been shown that for mass ratios $q = {M_2 \over M_1} \le 0.1$, the steady-state geometry of the disc in the binary 
frame is found to be quite different from that expected at higher mass ratios. For $q \sim 0.1$, the disc takes on the usual 
elliptical shape, with the major axis aligned perpendicular to the line of centres of the two stars. However, at smaller
mass ratios the elliptical gaseous orbits in the outer regions of the disc are rotated, with the major axis inclined in the 
direction of the secondary star. At $q = 0.01$, the major axis of these orbits is aligned almost parallel to the line of centres 
of the two stars. The rotation is more pronounced in hot, viscous discs, though there is no apparent correlation with the rate of mass
transfer from the donor star.

The key ingredient in determining the magnitude of the rotation angle is the maximum radius of the accretion disc and, in 
particular, the distance from the inner Lagrangian point to the disc edge. The disc occupies a large fraction of the Roche lobe at 
small mass ratios and the $L_1$-disc distance can be small. Furthermore, a hot disc spreads further under the action of
viscosity than a cool disc.

\subsection{Tidal truncation radii}
\begin{figure}
\hspace{0.5cm} \psfig{file=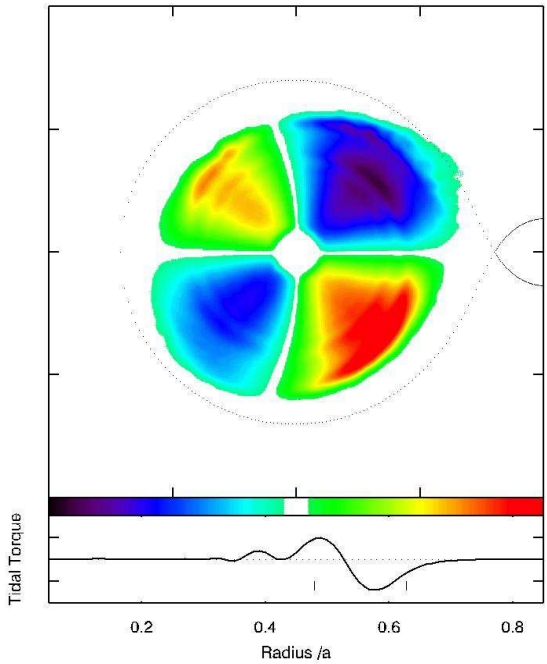,width=8cm}
\caption{Tidal torques on an accretion disc for $q = 0.03$, showing the alternating quadrants of positive (green to red) and 
negative (black to blue) torques. The lower panel is an azimuthally-averaged radial torque profile (with arbitrary units
of torque) and shows that the averaged contribution is only significant in the outer regions of the disc. The theoretical 
positions of the 3:1 and 2:1 resonances are marked by short vertical lines.}
\label{torque}
\end{figure}
It is clear that there is some value in being able to make a reliable estimate of the maximum radius of an accretion disc.
In the absence of any other significant torques (for example, a strong stellar magnetic field anchored on the primary star),
the disc is truncated by the tidal interaction with the secondary star, somewhere inside the Roche lobe boundary. The 
tidal/viscous truncation radius can only be determined numerically (Frank, King \& Raine 2002), so a fully three-dimensional hydrodynamical
model such as the one discussed here provides an ideal tool. Figure \ref{torque} shows the tidal torque field across the 
accretion disc for a binary system with $q = 0.03$. The alternating quadrants of positive and negative torque described by 
Frank, King \& Raine (2002) are clearly visible and the benefit of modern computer power allows the resolution of a wealth of 
detail associated with the spiral structure in the disc. The lower panel of Figure \ref{torque} shows the radial profile
of azimuthally-averaged torque. The net torque contribution is negligible in the inner parts of the disc, but very significant
where the spiral arms are most prominent in the outer parts. This result supports the findings of Blondin (2000), who 
found that the angular momentum transport in a disc due to tidal torques was not significant in the inner regions, but large 
near the disc edge (driving an effective viscosity parameter $\alpha \sim 0.1$).

Frank, King \& Raine (2002) associate the tidal radius of the disc with the point at which the torque first changes sign; in the case of Figure 
\ref{torque} this is near $R = 0.52a$. However, for discs which deviate from azimuthal
symmetry, it is clear that a single radius does not describe the geometry of the disc fully. This certainly will be the case 
for discs in binaries with small mass ratios. In the example shown here, for $q = 0.03$, the disc radius varies with azimuth
from $R =  0.54a$ to $R = 0.68a$. It remains true that the estimate
\begin{equation}
{R_{\rm tide} \over a} \simeq 0.9R_1 
\end{equation}
(Frank, King \& Raine (2002)), where $R_1$ is the radius of the primary Roche lobe, is representative of the {\em average} radius of the disc. 
For $q = 0.03$, this gives $R_{\rm tide} = 0.6{\rm a}$, with $\sim 10\%$ variation over the full range of radii. 

\subsection{Application to GP Com and other ultra-compact binaries}
The predicted shape of the accretion disc at low mass ratio - elliptical and fixed in the binary frame at an angle to the
line-of-centres - has been observed in the AM CVn system GP Com, which has an orbital period of 46 minutes and has a 
mass ratio $q \sim 0.02$ (Marsh 1999). Morales-Rueda et al. (2003) found that the He I emission in 
Doppler tomograms of GP Com had elliptical rather than circular symmetry. Fits to equal flux contours were found to be
ellipses with a maximum eccentricity $e \sim 0.6$. The ellipses are not aligned perpendicular to the line-of-centres 
(Figure 7 of Morales-Rueda et al. 2003). Morales-Rueda et al. (2003) 
have suggested that the discrepancy between their observation and the predictions of the restricted three-body model may
indicate an error in the ephemeris of GP Com. However, the simulations described in this paper suggest that no such problem
exists. Although the observed geometry of the disc disagrees with the inviscid three-body model, it is entirely consistent
with the predictions of a three-dimensional, viscous hydrodynamic model which includes the effects of mass transfer.
\begin{figure}
\psfig{file=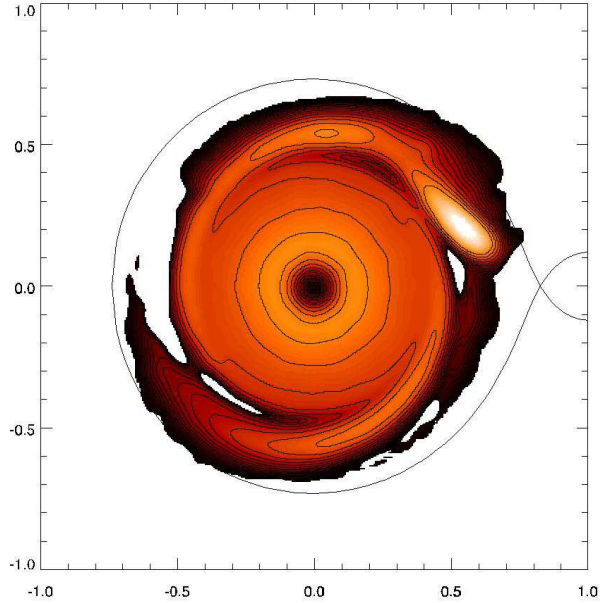,width=8cm}
\caption{Contours of energy dissipation rate per unit area in a disc with $q = 0.02$. The contours of constant dissipation are 
circular in the inner parts of the disc but elliptical in the outer parts. The stream-impact hot-spot is clearly visible. The 
colour scale is logarithmic, running from $10^5 \,{\rm erg\,s^{-1}cm^{-2}}$ (darkest shade) to 
$10^{11}\,{\rm erg\,s^{-1}cm^{-2}}$ (lightest shade).}
\label{tomo}
\end{figure}

Figure \ref{tomo} shows the contours of total energy dissipation rate per unit area in a simulation of a disc in a binary with
$q = 0.02$. The emission contours from the inner regions of the disc are circular, while those from the cooler, outer disc
are broadly elliptical, with some sub-structure associated with the hot spot and spiral arms. The angle of these elliptical 
contours from the normal to the line-of-centres of the two stars is about 
40 degrees. The ratio of the semi-minor to semi-major axes of the outer orbits is $0.83$, giving an eccentricity $e = 0.55$. These 
results compare favourably with the geometry of GP Com inferred from the Doppler maps of Morales-Rueda et al. (2003).

Since the angle is expected to vary for different mass ratios, its measurement may be a useful diagnostic for the
determination of mass ratios in very short period binaries. The angle does not specify $q$ absolutely uniquely; for example,
it can be seen from Figure \ref{bigone} that the tilt angle for $q = 0.03$ and $c_0 = 0.025$ (run 2) is similar to that for 
$q = 0.1$ and $c_0 = 0.05$ (run 9). However, if more information is available, or if reasonable assumptions can be made
regarding the temperature of the disc, then the angle can be associated with a narrow range of possible mass ratios. 
Furthermore, for $q \ale 0.01$, the disc geometry is unique and fairly independent of temperature.

\section{Concluding Remarks}
The use of measurements of the rotation angle of an accretion disc to place constraints on the binary mass ratio should represent
a useful diagnostic that can be used hand-in-hand with spectroscopic determinations. Further Doppler tomography of
ultra-compact systems is required to verify the predictions of the hydrodynamical models discussed here: the model will
stand or fall by these new observations. It will be particularly interesting to see if the asymmetrical disc geometry 
predicted for discs in binaries with $q \sim 0.01$ is found observationally. The smallest ultra-compact binary mass ratio 
claimed so far is $q = 0.0125$ for CE-315 (Deloye, Bildsten \& Nelemans (2005), though Ruiz et al. (2001) have cited $q = 0.022$). No 
Doppler maps have appeared in the literature thus far for this system.

The independence of disc geometry on mass injection rate is slightly surprising and this should be a focus of future
theoretical work. The thermal properties of the gas in the stream-impact region are known to affect the extent to which
the stream can overflow the disc edge (Armitage \& Livio 1998). It is conceivable that the inclusion of full radiative transfer in the 
hot-spot region will lead to discs with slightly different radii. Further theoretical work on the behaviour of the tidal instability at these
very low mass ratios is also desirable. It is of particular importance to find out  whether the instability is suppressed 
completely below a critical threshold value of $q$. While we know that the resonance can be identified in an inviscid 3-body model
down to mass ratios far smaller than those discussed here, we do not know whether this will translate into a viable instability in a
viscous disc in a binary with $q \sim 0.01$. It can be seen from Figures \ref{bigone} and \ref{bigtwo} that the disc at $q$ = 0.01 already
fills most of the Roche lobe long before any measurable precession is detected. It is not clear at all how any subsequent precession could
be maintained: inevitably there would be significant mass loss across the Roche lobe boundary. 

The steady state geometry of discs in extreme mass ratio systems predicted by this work is an example of a physical situation
in which the inviscid, restricted three-body model breaks down when applied to accretion discs. Failures of the model are 
surprisingly rare. The last non-intersecting three-body orbit remains a good estimate of the tidal truncation radius for 
$q > 0.1$, while the critical upper limit for tidal instability at $q = 0.25$ in the three-body problem agrees well with both 
observations of cataclysmic variables and previous numerical calculations. However, at $q < 0.1$, the disc can fill a very 
large fraction of the Roche lobe. The presence of a non-zero viscosity coupled with the action of the tides leads to a 
strong, asymmetric torque on the gas, as shown in Figure \ref{torque}. The asymmetry is most pronounced for the largest 
orbits, and it is this that leads to the rotation of the orbits away from the perpendicular to the line-of-centres.

\section*{Acknowledgements}
This work was instigated during a visit to the Centre for Astrophysics and Supercomputing at Swinburne University of 
Technology in Melbourne in late 2005. I would like to thank them, and in particular James Murray for their hospitality 
during my stay. I am also grateful to Gordon Ogilvie for a helpful discussion about the characteristics of the tidal instability
and to the anonymous referee for several constructive comments.

\section*{Bibliography}

Armitage P., Livio M., 1998, ApJ, 493, 898\\
Blondin J.M., 2000, New Astron., 5, 53\\
Deloye C.J., Bildsten L., Nelemans G., 2005, ApJ, 624, 934\\
Frank J., King A.R., Raine D.J., 2002, Accretion Power in Astrophysics, 3rd edn.\\ 
\indent Cambridge University Press\\
Goodchild S., Ogilvie G., 2006, MNRAS, 368, 1123\\
H\'enon M., 1965, Ann. d'Ap, 28, 992\\
Hirose M., Osaki Y., 1990, PASJ, 42, 135\\
Kunze S., Speith R., 2005, The Astrophysics of Cataclysmic Variables and Related\\
\indent  Objects, eds. Hameury J.-M., Lasota J.-P., ASP Conference Series, 330, 389\\
Kunze S., Speith R., Riffert H., 1997, MNRAS, 289, 889\\
Lubow S.H., 1991, ApJ, 381, 259\\
Lubow S.H., 1994, ApJ, 432, 224\\
Lubow S.H., Shu F.H., 1975, ApJ, 198, 383\\
Lyubarskij Y.E., Postnov K.A., Prokhorov M.E., 1994, MNRAS, 266, 583\\
Marsh T.R., MNRAS, 304, 443\\
Matthews O.M., Wheatley P.J., Wynn G.A., Truss M.R., 2006, MNRAS, 372, 1593\\
Monaghan J.J., 1992, ARA\&A, 30, 543\\
Morales-Rueda L., Marsh T.R., Steeghs D., Unda-Sanzana E., Wood J.H., North R.C.,\\ 
\indent 2003, A\&A, 405, 249\\
Murray J.R., 1996, MNRAS, 279, 402\\
Murray J.R., 1998, MNRAS, 297, 323\\
Murray J.R., Warner B., Wickramasinghe D.T., 2000, MNRAS, 279,402\\
Ogilvie G.I., 2001, MNRAS, 325, 231\\
Osaki Y., Meyer F., 2002, A\&A, 383, 574\\
Paczy\'nski B., 1977, ApJ, 216, 822\\
Piotrowski S.L., Ziolkowski K., 1970, Ap\&SS, 8, 66\\
Ruiz M.T., Rojo P.M., Garay G., Maza J., 2001, ApJ, 552, 679\\
Simpson J.C., Wood M.A., 1998, ApJ, 506, 360\\
Truss M.R., 2005, MNRAS, 356, 1471\\
Truss M.R., Murray J.R., Wynn G.A., 2001, MNRAS, 324, L1\\
Whitehurst R., 1988, MNRAS, 233, 705\\
Whitehurst R., King A.R., 1991, MNRAS, 249, 25\\

\label{lastpage}

\end{document}